\documentclass[conference]{IEEEtran}
\IEEEoverridecommandlockouts
\usepackage{cite}
\usepackage{amsmath,amssymb,amsfonts}
\usepackage{amsthm}
\usepackage{tabulary}
\usepackage{url}
\usepackage{booktabs}
\usepackage{nccmath}
\usepackage{mathtools}

\usepackage{algorithmic}
\usepackage{graphicx}
\usepackage{float}
\usepackage{textcomp}
\usepackage{xcolor}

\usepackage{setspace} 
\usepackage{array} 
\usepackage{paralist} 
\usepackage{verbatim} 
\usepackage{subfig} 
\usepackage{todonotes}

\def\BibTeX{{\rm B\kern-.05em{\sc i\kern-.025em b}\kern-.08em
    T\kern-.1667em\lower.7ex\hbox{E}\kern-.125emX}}

\newcounter{l2}
\newcommand{\balphlist}{\begin{list}{(\alph{l2})}{\usecounter{l2}}}
\newcommand{\barablist}{\begin{list}{\arabic{l1}}{\usecounter{theorem}}}

\usepackage{pgfplots}
\usetikzlibrary{patterns}
 \usetikzlibrary{
        pgfplots.colorbrewer,
    }

\pgfplotsset{compat=1.16} 
\pgfplotsset{width=7.5cm,compat=1.5.1}

\begin{document}

\IEEEoverridecommandlockouts
\IEEEpubid{\text{\begin{minipage}[t]{\textwidth}\ \\[10pt]
       {This article has been accepted for presentation at the 2021 IEEE International Conference on Communications, Control, and Computing Technologies for Smart Grids (SmartGridComm). DOI: 10.1109/SmartGridComm51999.2021.9632293 
        \copyright 2021 IEEE. Personal use of this material is permitted. Permission from IEEE must be obtained for all other uses, in any current or future media, including reprinting/republishing this material for advertising or promotional purposes, creating new collective works, for resale or redistribution to servers or lists, or reuse of any copyrighted component of this work in other works.}
\end{minipage}}} 

\newcounter{foo}
\newcounter{boo}
\newcounter{res}
\newtheorem{theorem}[foo]{Theorem}
\newtheorem{lemma}[foo]{Lemma}
\newtheorem{definition}[boo]{Definition}
\newtheorem{remark}[foo]{Remark}
\newtheorem{corollary}[foo]{Corollary}
\newtheorem{result}[res]{Result}
\renewcommand{\thetheorem}{\arabic{foo}}
\renewcommand*{\proofname}{Proof Sketch}

\newtheorem*{discussion}{Discussion}

\title{Asset Participation and Aggregation in Incentive-Based Demand Response Programs
\thanks{The authors are with the Department of Electrical Engineering and Computer Sciences, University of California, Berkeley}
}
\author{Utkarsha Agwan, Costas J. Spanos, and Kameshwar Poolla}

\maketitle

\begin{abstract}
In order to manage peak-grid events, utilities run incentive-based demand response (DR) programs in which they offer an incentive to assets who promise to curtail power consumption, and impose penalties if they fail to do so. We develop a probabilistic model for the curtailment capability of these assets, and use it to derive analytic expressions for the optimal participation (i.e., promised curtailment) and profitability from the DR asset perspective. We also investigate the effects of risk-aversion and curtailment uncertainty on both promised curtailment and profit. We use the probabilistic model to evaluate the benefits of forming asset aggregations for participation in DR programs, and develop a numerical test to estimate asset complementarity. We illustrate our results using load data from commercial office buildings.
\end{abstract}


\section{Introduction}

Demand response (DR) programs are tools to modulate the demand for electricity in a wide variety of situations. For example, at certain times such as mid afternoons on hot summer days, the procurement of additional electricity to meet demand peaks is expensive. At these times, it is more cost effective to induce a reduction in demand than to procure an increase of supply to maintain power balance \cite{siano2014demand}. Another scenario is a grid with significant renewable generation assets. Here DR promises to be a superior balancing resource when compared with conventional gas turbines in the metrics of cost and emissions. Realizing its potential the 205 Energy Policy Act promotes trading DR assets in organized wholesale energy markets. FERC Order 745 mandates that DR assets be compensated for load reduction on par with generation \cite{ferc745}.

Participation in DR programs requires assets to accurately estimate their curtailable load and stake money (as penalties or as forgone payments) on their ability to offer the promised load reduction. The curtailable load is a random variable and depends on the operating status of the asset at the time of the DR event, which makes it difficult to determine the optimal load reduction that can be promised. Asset aggregations can benefit participants by spreading the risk of default on the promised curtailment, and choosing aggregation members optimally can offer a significant marginal benefit to the participants. 

 \subsection{Demand Response Programs}
 
There are two flavors of DR programs: \textit{direct load control}, where utilities can modulate loads at will subject to certain contractual limits, and \textit{indirect load control}, where agents are incentivized to yield their discretionary electricity consumption through payments \cite{siano2014demand}. This paper focuses on the latter.

We explore the situation where the utility enters into bilateral contracts with a collection of agents. These agents might be buildings with flexible electricity demand, or aggregators that manage a collection of DR assets. An agent which enters into a contract of size $C$ kWh receives a payment of $\pi_{\text{r}} C$ from the utility for their obligated reduction in demand, where $\pi_{\text{r}} $ is the incentive rate set by the utility. In the event the agent is unable to deliver the contractual demand reduction $C$, it pays a penalty proportional to the shortfall. There are often contractual riders that limit the frequency with which DR assets are called on in response to a DR event.



%


 \subsection{Contributions}

In this paper, we consider the perspective of DR assets which aim to maximize their net payments from participating in the DR program while capping their lost utility (i.e., \textit{disutility}) from reduced consumption. The assets will place a cap on the contract size allowable based on how frequently they expect a DR event to be called, and then optimize their contract size. We consider a time-varying DR contract, i.e. a different curtailment size is offered in each time window (e.g. 1 hour). For this program model, we develop analytic expressions for

\begin{enumerate}
\item The optimal contract and profitability of the asset,
\item The effect of risk aversion and curtailment uncertainty on contract size and profit,
\item The value of aggregations and the effect on contract size and profitability, and
\item A metric for assessing asset complementarity.
\end{enumerate}

\subsection{Related Work}

    A rich body of work exists on modeling demand response contracts, and optimizing asset participation based on disutility models or by assuming fully controllable flexible loads. However, the uncertain nature of realized curtailment has not been investigated as deeply.  \cite{campaigne2016firming} develop a probabilistic model for DR curtailment and use it to formulate an aggregator's day ahead offer in a wholesale market.  \cite{jung2014data} use lookup tables of historical consumption data to estimate flexibility, and then model the total power consumption as a Gaussian variable to determine DR capacity. \cite{bitar2012bringing} model the optimal offering for uncertain wind producers by using a probabilistic framework.

    Demand response aggregations exist in practice, with several companies aggregating industrial or home-thermostat loads to provide services to utilities. Researchers have investigated the problem of optimizing such aggregations, especially when coupled with uncertain renewable generation. \cite{henriquez2017participation} model optimal portfolios for DR aggregators participating in day ahead and real time markets, and model market price uncertainty but assume that flexible load is controllable with some disutility cost.     \cite{parvania2013optimal} optimize DR asset aggregations but do not model the uncertain nature of realized curtailments. \cite{ghose2019risk} work on assessing the financial risk of microgrid aggregators with DR resources using the CVaR metric.
    
    
    Estimating DR capability is an essential part of the problem of optimizing DR participation. \cite{wang2020smart} forecast aggregated smart-thermostat loads and use it to estimate demand flexibility from a DR aggregator's perspective. \cite{ali2014demand} estimate that around 60\% of residential HVAC load is curtailable with a relatively small effect on thermal comfort. DR baselining methods such as CalTrack \cite{caltrack} rely on using the temperature and time-of-day dependence of electric load to forecast future usage. 

\section*{Notation}

\begin{tabular}{ c | l }
$C$ & Size of contract signed by the asset $(kWh)$ \\
$q$ &  Curtailment capability (random variable) $(kWh)$\\
$f(q)$ & Probability density function of $q$ \\
$F(q)$ & Cumulative distribution function of $q$ \\
$\pi_{\text{e}}$ & Retail tariff paid by the building \\
& for energy consumed $(\$/kWh)$\\
$p$ & \emph{Ex-ante} probability of DR event  $(h^{-1})$\\
$\pi_{\text{r}}$ & Incentive paid by utility to participate   \\
&  in DR program, per unit of \\
& curtailment contracted $(\$/kWh)$ \\
$\pi_{\text{p}}$ & Penalty imposed on shortfall 
 in curtailment \\
 &during a demand response event $(\$/kWh)$ \\
\end{tabular}


\section{Problem Formulation}
Consider a forward window for the delivery of demand response services, e.g. Tuesday 3-4 pm next week. The retail price of electricity during this window is $\pi_{\text{e}}$. Let $p$ denote the probability of a DR event in this window which can be estimated \emph{ex ante} using historical data and exogenous forecasts of temperature and demand, and utilities often publish this information on public-facing dashboards \cite{coned-dr}. An agent with DR assets (ex: a building with flexible load) can provide demand reduction during this window. The agent's available demand reduction $q$ in this forward window is a random variable with density function $f(\cdot)$ and cumulative density function $F(\cdot)$. This reduction $q$ depends on realizations of underlying random processes such as temperature, total load, occupancy and other variables which determine the density functions $f(\cdot), F(\cdot)$. The agent promises to reduce its demand by $C$ kWh to the utility codified through a bilateral contract which may have additional riders, and the utility pays the agent $\pi_{\text{r}} C$ in advance for this promise. The reward price $\pi_{\text{r}}$ can depend on the delivery window, and is set by the utility and published \emph{ex ante}. Because the available demand reduction $q$ is a random variable, the agent may not be able to meet its demand reduction promise of $C$ kWh. The utility imposes an \emph{ex post} penalty  $\pi_{\text{p}} (C - q)^+$ on the shortfall $(C - q)^+ = \max(0,C-q)$, to be paid by the agent.



\subsection{Agent Decisions}
During the DR event, the agent has no incentive to curtail its demand beyond the contracted value $C$, as it receives no additional payment for doing so. The realized curtailment is
\begin{equation}\label{eqn:x}
 x = \min (C, q),
\end{equation}
where $q$ is the agent's available demand reduction capability over the DR window. The profit of the DR agent is a random variable, and depends on whether a DR event occurs. If a DR event is called, the realized profit is 
\begin{equation}\label{eq:profit-DRevent}
    \pi_{\text{r}} C - \pi_{\text{p}} (C-x) + \pi_{\text{e}} x,
\end{equation}
and the realized profit in the absence of a DR event is 
\begin{equation}\label{eqn:random-profit-noDR}
    \pi_{\text{r}} C.
\end{equation}
Note that the agent receives a payment for its contract $C$ irrespective of whether or not a DR event occurs. The three components of the agent profit are a payment $\pi_{\text{r}} C$ for the contract, a penalty $\pi_{\text{p}} (C-x)$ on the shortfall, and a surplus $\pi_{\text{e}} x$ from avoided electricity use. The agent's profit is a random variable, and it will seek to maximize its expected profit


\begin{equation}\label{eq:agent-profit}
\begin{split}
J^{\text{a}} = \pi_{\text{r}} C + p \left\{ - \pi_{\text{p}} \textstyle\int\limits_0^C (C-q) f(q) dx  \right. \\
\left. +  \pi_{\text{e}} \left[ \textstyle\int\limits_0^C q f(q) dq + \textstyle\int\limits_C^{\infty} C f(q) dq \right] \right\}.
\end{split}
\end{equation}



\begin{remark}
In reducing their electricity consumption during a DR event, the agent will suffer some disutility. For example, the occupants of a building that reduces its electricity consumption from baseline may suffer the inconvenience of elevated temperatures. Building managers may not be able to assign a monetary value to this disutility. A simple approach is to regard the agent's available demand reduction as yielding their discretionary electricity consumption. We can therefore simply bound the contract size by $C^{\text{max}}$ to reflect the maximum available discretionary electricity consumption, i.e.
\begin{equation} \label{eqn:Cmax}
q \leq C \leq C^{\text{max}}.
\end{equation}
Another approach is to explicitly introduce a disutility term. Agents will cede some of their baseline consumption only if the marginal disutility is lower than the marginal profit. While these can be readily incorporated into our formulation, we will not do so to keep our exposition simple.

\end{remark}

\begin{remark}
Agents providing DR may be risk averse. For example, flexible buildings may wish to maximize their expected profit while bounding their worst case loss. Risk aversion can be handled through metrics such as CVaR, which is a quantification of the tail risk.
\begin{equation} \label{eq:agent-cvar}
\begin{split}
\text{CVaR} = \pi_{\text{r}} C + \frac{p}{1-\hat{c}}  \int\limits_0^{F^{-1}(1-\hat{c})} \left[ \pi_{\text{e}}  q  - \pi_{\text{p}} (C-q) \right] f(q) dq
\end{split}
\end{equation}
\end{remark}

The risk averse agent will then optimize a combination of its expected profit and the risk assessment measure, weighted by a risk aversion factor $\alpha$ as: 
\begin{equation}\label{eq:agent-opt}
    \begin{split}
        \max\limits_C & \: \:  J^{\text{a}} + \alpha \text{CVaR}(C)\\
\text{s.t.}  &\: \: C \leq C^{\text{max}}
    \end{split}
\end{equation}


\subsection{Demand Response Contracts}
    Utilities are mandated to reliably and adequately procure electricity to meet the random needs of their customers, and they conduct demand forecasts to drive their purchase decisions in the day ahead market.  They subsequently purchase additional electricity in the real time market for fine balance of supply and demand, at prices that may vary widely.
    Utilities absorb price volatility in these two settlement markets, and resell procured electricity to their customers at a retail tariff $\pi_{\text{e}}$ which may be fixed or have time of use structure. At times when the aggregate demand is substantial enough to strain generation resources, the price of electricity will be large enough that it will be in the utility's financial interest to displace additional procurement with demand reduction through DR programs \cite{siano2014demand}. The utility might seek demand reduction for other reasons, such as environmental concerns around scheduling high emissions peaker plants, available capacity, and transmission constraints. There are two reasons why a utility could choose an incentive based program instead of a real time DR market:


    
    
    \begin{enumerate}
    
    \item \emph{Price risk.} FERC regulations dictate that DR assets be compensated for their curtailment on par with market rates. Contracts insulate the utility from high real time prices by committing DR assets to \emph{ex ante} compensation. 
    \item \emph{Quantity risk.} The curtailment realized from DR assets is uncertain. Contract mechanisms can reduce this uncertainty through a well designed penalty structure. This is more effective than relying on very accurate curtailment forecasts which is necessary in organized markets.
\end{enumerate}

\section{Optimal Agent Decisions}
Recall that the expected profit of the DR agent is given in \eqref{eq:agent-profit}, and
the agent's CVaR risk aversion metric is given in \eqref{eq:agent-cvar}.
We consider optimal agent decisions in a single DR window. Assume $\pi_{\text{r}} - p \pi_{\text{p}} < 0$. 


\begin{result} \label{res:contract-profit} \textit{Optimal Contract and Profitability} \begin{enumerate}
\item The optimal contract size that maximizes the agent's risk averse expected profit (see Equation \ref{eq:agent-opt}) is given by
\begin{equation}\label{eq:agent-contract}
 C^* = F^{-1} \left[ \frac{\pi_{\text{r}} +  p \pi_{\text{e}} + \alpha (\pi_{\text{r}} - p  \pi_{\text{p}})}{p (\pi_{\text{p}}  + \pi_{\text{e}})} \right]
\end{equation}
\item The corresponding expected profit is
\begin{equation} \label{eq:agent-optimal-profit}
\begin{split}
J^{\text{a}*} &= p (\pi_{\text{p}} + \pi_{\text{e}}) \int_0^{C^*}  q f(q) dq- \alpha  (\pi_{\text{r}} - p \pi_{\text{p}}) C^*
\end{split}
\end{equation}
\item Risk aversion, i.e. increasing $\alpha$ decreases the expected profit and the optimal contract size, but reduces worst case losses.
\end{enumerate}
\end{result}
\begin{proof}
The agent's optimization objective \eqref{eq:agent-opt} is concave in the contract size $C$, which can be proven with the second derivative test. Then the optimal contract $C^*$ can be calculated from the zero of the second derivative of the objective and clipping it to $[0,C^{\text{max}}]$. The expression for profit can be calculated by substituting $C^*$ in \eqref{eq:agent-profit}.
\end{proof}
The usefulness of this result hinges on being able to calculate the quantile in \eqref{eq:agent-contract}. While the distribution of curtailment capability at any time may not be readily available, it can be estimated using a combination of historical data, operational information and forecasts of causal variables like weather. We present one potential method to build a data-driven distribution based on heuristics founded in literature and practice in Section \ref{sec:examples}. 

\begin{remark}
$\pi_{\text{r}} - p \pi_{\text{p}}$ is the expected profitability of participating in the DR program without any underlying curtailment capability, i.e. without any backing asset. The utility will set reward and penalty such that $\pi_{\text{r}} - p \pi_{\text{p}} < 0$; otherwise, it will encourage unwanted behavior, such as bidding without any DR asset capability.
\end{remark}

The optimal contract $C^*$ and corresponding profit depend intimately on the distribution of curtailment capability $f(\cdot)$. The standard deviation $\sigma$ of the curtailment distribution is a measure of the uncertainty in its estimation. A smaller $\sigma$ signifies a tighter estimate of future capability, and reduces the risk of shortfall. Under a mild symmetry assumption on $f$, we have

\begin{result}\label{res:variance}\textit{Effect of Uncertainty in Future DR Capability}
\begin{enumerate}
\item \textit{Contract:} Suppose $f$ is symmetric about the mean $\mu$, with standard deviation $\sigma$, then
\begin{equation}
C^* = \mu + \gamma \sigma 
\end{equation}
where $\gamma$ is a function of $p,\pi_{\text{r}} , \pi_{\text{p}}, \pi_{\text{e}}, \alpha$ derived from the expression for optimal contract (\ref{eq:agent-contract}). If $\gamma < 0$, i.e. contract is lower than $\mu$, then decreasing  $\sigma $ leads to a higher contract. If $\gamma > 0$, decreasing $\sigma$ leads to lower contract. A decrease in $\sigma$ pushes the contract size closer to $\mu$. 


\item  \textit{Profit:}
The coefficient of $\sigma$ is negative up until a certain positive $\gamma$ value, i.e. until $\gamma = \hat{\gamma}$, and positive after that. For $\gamma < \hat{\gamma}$, an increase in uncertainty (i.e., increase in $\sigma$) causes a decrease in expected profit.


\end{enumerate}
\end{result}
The symmetry assumption is a reasonable model, particularly where the curtailment capability has a `base' value with some error due to operational variability. In fact, the data we use in Section \ref{sec:examples} shows symmetry. 

This result could be useful in evaluating the benefits of lowering uncertainty. For example: paying for more accurate weather forecasts, or paying to install submetering equipment in buildings could add value by leading to tighter estimations of future DR capability. Knowing the value of reduced uncertainty (i.e., lower $\sigma$) in terms of increased DR profit can help in making a decision on the cost-benefit tradeoff. Also, as the DR event window approaches, the estimation of DR capability might improve (e.g. predicting DR capability one month in advance vs. a few hours in advance). Our results will be useful for these near-term predictions as well, which might have lower uncertainty.

\section{Optimal Aggregations and Decisions}
DR assets can be aggregated to provide DR as a single entity, sharing the burden of curtailing load and reducing their individual risk. We will not explore how DR assets can be induced to form such aggregations; instead, we focus on optimal decisions of the aggregation as a joint entity.

For the results presented in this section, we make the following assumption: DR agents' curtailments are similarly distributed (i.e. from the same family of probability distributions), but with different mean and variance parameters $\mu, \sigma$. Further, the sum of agents' curtailments is distributed similarly as well (ex: normal distribution), and the distribution $f_k(\cdot)$ of the curtailment capability of asset $k$ is completely characterized by its mean and variance. From Result \ref{res:variance}, the optimal contract for asset $k$ is
\begin{equation}
C_k^* = \mu_k + \gamma \sigma_k
\end{equation}
where $\gamma$ is completely determined by utility and DR program rates. The aggregate DR asset curtailment capability is
\begin{equation}\label{eq:aggregation-curtailment}
\begin{split}
q_{\text{ag}} = \sum\limits_k q_k 
\end{split}
\end{equation}
The optimal contract for the aggregation differs from the sum of contracts for the individual participants due to the different variance.
\begin{equation} \label{eq:aggregation-contract} 
\begin{split}
C_{\text{ag}}^* &= \mu_{\text{ag}}  + \gamma \sigma_{\text{ag}} = \sum_k \mu_k + \gamma \sigma_{\text{ag}}\\
\sum_k C_k^* &= \sum_k \mu_k + \gamma \sum_k \sigma_k\\
\end{split}
\end{equation}

\begin{result} \textit{Aggregation Contract Size and Profitability} \label{res:aggregation-contract}
\begin{enumerate}

\item The optimal contract for the aggregation is
\begin{equation}
\begin{split}
 C^*_{\text{ag}} = F^{-1}_{\text{ag}} \left[ \frac{\pi_{\text{r}} +  p \pi_{\text{e}} + \alpha_{\text{ag}} (\pi_{\text{r}} - p \pi_{\text{p}})}{p (\pi_{\text{p}}  + \pi_{\text{e}})} \right]
\end{split}
\end{equation}
where $\alpha_{\text{ag}}$ is the risk aversion factor for the aggregation, and $F_{\text{ag}}(\cdot)$ is the cumulative density function for the aggregation's curtailment capability $q_{\text{ag}}$.

\item The relation between the aggregation's contract size and the aggregate contract of individual participants is
\begin{equation}\label{eq:agg-contract-comparison}
\begin{split}
C_{\text{ag}}^* & \left\{\begin{array}{lr}  
< \sum\limits_b C_b^* \: \: \:  \text{if} \: \: \:  \gamma > 0\\
> \sum\limits_b C_b^* \: \: \:  \text{if} \: \: \: \gamma < 0
\end{array}\right.
\end{split}
\end{equation}

\item At the optimal contract, total increase in profit for an aggregation of assets with similar $\alpha$ is
\thickmuskip=5mu plus 3mu minus 5mu
\medmuskip=4mu plus 2mu minus 4mu
\begin{equation}
\begin{split}
\Delta J^{\text{a}} &=  - \alpha(\pi_{\text{r}} - p \pi_{\text{p}}) (C_{\text{ag}} - \sum_k C_k) \\
& +p (\pi_{\text{p}} +  \pi_{\text{e}}) \left[ \textstyle\int\limits_0^{C_{\text{ag}}} q f(q) dq  - \textstyle\sum_k  \int\limits_0^{C_k} q_k f_k(q_k) dq_k  \right]
\end{split}
\end{equation}
\end{enumerate}
\end{result}
We examine the profit differential for assets whose curtailments are normally distributed, i.e.  $q_k \sim \mathcal{N} (\mu_k , \sigma_k)$; $C_k^* = \mu_k + \gamma \sigma_k$. For a normal distribution, the quantile function is given by 
\begin{equation*}
\begin{split}
    F^{-1}(p) = \mu + \sigma [\sqrt{2}\text{erf}^{-1}(2p-1)] = \mu + \sigma \gamma 
\end{split}
\end{equation*}
which gives us the expression 
\begin{equation}\label{eqn:del-J}
    \begin{split}
        \Delta J^{\text{a}} = -p (\pi_{\text{p}} +  \pi_{\text{e}}) (N_k - 1)\frac{1}{2}\left[ \text{erf} \left( \textstyle\frac{\gamma}{\sqrt{2}}\right) +1\right]\\
        + \left( \textstyle\sum_k \sigma_k - \sigma_{\text{ag}}\right) \left\{ p (\pi_{\text{p}} +  \pi_{\text{e}}) \textstyle\frac{e^{-\frac{\gamma^2}{2}}}{\sqrt{2 \pi}} + \alpha(\pi_{\text{r}} - p  \pi_{\text{p}}) \gamma \right\}
    \end{split}
\end{equation}






We define an optimal aggregation as one which maximizes the increase in \emph{social welfare}, i.e. leads to the greatest increase in profit for its participants. In the expression for profit increase \eqref{eqn:del-J}, we can see that the mean of the distributions of curtailment capability doesn't affect the marginal benefit, rather the variability does. If the random variables $q_k$ (the DR capability) of two assets are highly correlated, they will have a lower marginal benefit in aggregating as compared to two uncorrelated assets. The intuition behind this is that uncorrelated variability de-risks the contract commitment and reduces the total variability of the sum of curtailment capacities.






\begin{definition}\label{def:complementarity}\textit{Complementarity Test}\\
To evaluate DR assets $k \in \mathcal{N}$ that aim to form an aggregation, the metric 
\begin{equation}\label{eqn:complementarity}
    \Delta \sigma = \textstyle\sum\limits_{k \in \mathcal{N}} \sigma_k - \sigma_{ag}
\end{equation}
is an indicator of the marginal increase in profitability upon aggregation. Here $\sigma_k$ is the standard deviation of the distribution of curtailment capability for asset $k$, and $\sigma_{ag}$ is the standard deviation of the distribution of the sum of curtailment capabilities for all assets $k \in \mathcal{N}$. 
\end{definition}
$\sigma_{ag}$ is lower than the sum of individual $\sigma_k$, as it exploits the low correlations across different assets. The lower the correlation of two assets, the higher this difference, (and consequently the marginal increase in profit) will be. This test can be useful for third-party DR aggregators that want to optimally package or aggregate assets while signing curtailment contracts. While it may seem that the information needed to calculate this metric is difficult to obtain, it might be readily available to entities such as smart thermostat aggregators or battery management companies.

\section{Illustration}\label{sec:examples}

We now illustrate our theoretical models by applying them to demand response data. In our work, we tried to solve the problem of optimal participation if the DR capability distribution is known. However, forecasting demand response capability for assets is a complex challenge. Data from real demand response programs is scarce, and is often unavailable to researchers due to privacy concerns. To get around the problem of data unavailability, we present a few heuristics from literature that can be used to estimate DR capability from load data, and then apply those to real load data from commercial buildings to generate distributions that can feed into our model. 

\subsection{Method for estimating distributions $f(q), F(q)$} \label{sec:dist-method}
\begin{enumerate}
    \item We assume that some part of HVAC load is curtailable for 1) short time durations and 2) at a low event frequency without a significant impact on occupant comfort. This is backed up by existing work, e.g. \cite{wang2020smart}, \cite{ali2014demand}.
    
    \item We model the building load as a linear combination of sample load-shapes for different end uses from \cite{loadshape}, and are able to obtain a good fit. 
    
    \item We assume that historical disaggregated loads are a good indicator of future loads both in quantity and composition, which is consistent with DR baselining. We use CalTrack methods \cite{caltrack} to form buckets for historical data. We create buckets for each hour of the day (e.g. 1 pm - 2 pm) in each month, and split across weekday/weekend to build distributions $f(q)$ from which future curtailment capability is assumed to be sampled.
    
    
    \item We fit a normal distribution to each bucket in order to evaluate our complementarity test. Note that this is not required to assess the optimal contract or profitability, which can be obtained from a quantile of the bucket data. 
\end{enumerate}
The distribution shown in Fig. \ref{fig:contract-risk} is estimated using this method. Note that this is one of many possible methods of constructing the distributions, and other methods can be used that better suit the resource type or data availability. The assets we consider are a set of commercial buildings in the Los Angeles area. These buildings are primarily used as office space, and were metered as part of the Building Genome project \cite{Miller2017439}. The dataset is anonymized by using names for buildings, which we also use as labels in our illustrations. We model incentive payments and penalties as in PG\&E's Base Interruptible Program \cite{pge-bip} which calls DR events in peak grid conditions, and model a low event frequency (3 hour-long events per month). We model a non-zero event probability for each hour of the month, which may not be the case in reality. Utilities can modulate their incentive payments to reflect estimated event probability, but that perspective is outside the scope of this paper. 


\subsection{Optimal Contract and Profitability}

\begin{figure}
\begin{tikzpicture}
	\begin{axis}[
	    height = 5cm,
		xlabel= Hour of day,
		ylabel style={align=center},
		ylabel= Contract size and \\ total load (kWh),
		legend style={
                font=\footnotesize,
                at={(0.5,1)},
                anchor=south,
                column sep=1ex,
                draw = none},]
	\addplot[color=blue,mark=x] coordinates {
(1, 158.96445)
(2, 154.1830583)
(3, 151.8251333)
(4, 150.8426417)
(5, 142.851875)
(6, 142.1968833)
(7, 137.5464917)
(8, 134.0750833)
(9, 137.8739917)
(10, 167.6101917)
(11, 241.8198333)
(12, 283.0182833)
(13, 285.9657083)
(14, 298.1483917)
(15, 308.497125)
(16, 323.0377667)
(17, 319.2388583)
(18, 319.9593333)
(19, 324.0202333)
(20, 309.7416083)
(21, 297.6244083)
(22, 279.677875)
(23, 204.3547833)
(24, 167.151725)
	};
\addplot[color=orange,mark=x] coordinates {
(1, 22.86223383)
(2, 22.14892042)
(3, 22.03189872)
(4, 21.12142627)
(5, 19.82582016)
(6, 17.90853666)
(7, 15.86178404)
(8, 14.24890314)
(9, 14.89895587)
(10, 20.61804529)
(11, 34.12168738)
(12, 43.21877545)
(13, 43.9722531)
(14, 47.10979641)
(15, 49.71437563)
(16, 51.90272785)
(17, 51.71658547)
(18, 53.4714484)
(19, 54.32521688)
(20, 53.97244544)
(21, 53.76098367)
(22, 50.90543129)
(23, 34.22497402)
(24, 26.81752326)
	};
\legend{Total electric load, Optimal contract size}
	\end{axis}
\end{tikzpicture}
\caption{Optimal contract size and total electrical load over the course of a day for a single building}\label{fig:contract-load}
\end{figure}
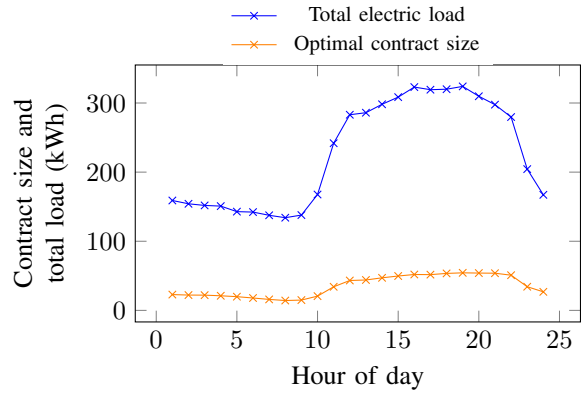

\begin{figure}
\includegraphics[width=\columnwidth]{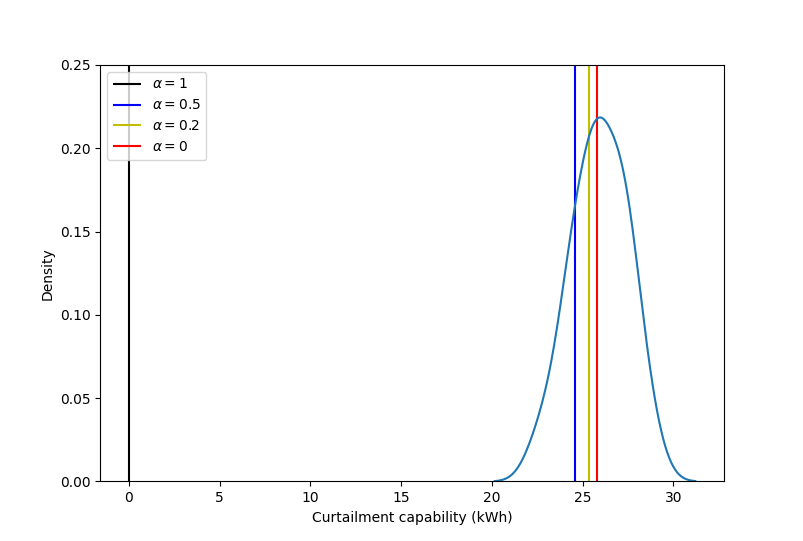}
\caption{Estimated probability density for a building at 1 pm on a weekday, and the dependence of optimal contract size on a varying risk aversion factor. Here, the optimal contract size ($C^*$) is indicated by a vertical line at $C^*$ kWh. } \label{fig:contract-risk}
\end{figure}

\begin{figure}
\begin{tikzpicture}
	\begin{axis}[
	    height = 5cm, 
		xlabel= Risk aversion factor ($\alpha$),
		ylabel style={align=center},
		ylabel=Profit from participation\\ in DR program over the \\ course of a month (\$)]
	\addplot[color=blue,mark=x] coordinates {
(0, 206.2414526)
(0.1, 203.0637818)
(0.2, 199.7819286)
(0.3, 196.316996)
(0.4, 192.5581229)
(0.5, 188.3282251)
(0.6, 183.2968384)
(0.7, 176.6414505)
(0.8, 165.196287)
(0.9, 0)
(1, 0)
	};
	\end{axis}
\end{tikzpicture}
\caption{Dependence of profit on risk aversion for a single building over the course of a month }\label{fig:profit-risk}
\end{figure}
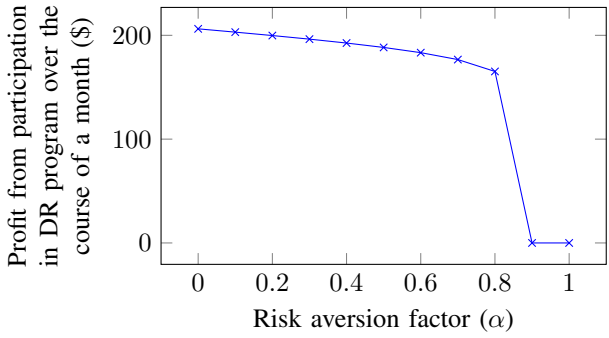

We first illustrate our results on optimal contract size and profitability in Result \ref{res:contract-profit}. In Fig. \ref{fig:contract-load} we illustrate the optimal contract sizes over the course of a day for a single building, and compare it with the total electric load over the same time. The optimal contract size is not a constant proportion of the total load, and depends on the proportion of flexible load. In Fig. \ref{fig:contract-risk} we illustrate the estimated distribution $f(q)$ (using the method outlined in Section \ref{sec:dist-method}) and optimal contract size for a single hour for a building. The optimal contract size is calculated as a quantile of the underlying data, as in \eqref{eq:agent-contract}. 
The optimal contract is not a simple function of the total load, as it depends on the proportion of flexible load as well as uncertainty in its estimation. 
In Fig. \ref{fig:contract-risk} and Fig. \ref{fig:profit-risk} we illustrate the effect of different levels of risk aversion (i.e., $\alpha$) on optimal contract and expected profit, respectively. At a certain $\alpha$ threshold, the quantile in \eqref{eq:agent-contract} becomes zero, and the optimal contract becomes the lowest curtailment that can be guaranteed ($C^* = 0$). This leads to a flattening out of the profit curve as seen in Fig. \ref{fig:profit-risk}.

\subsection{Optimal Aggregation, or Complementary Assets}

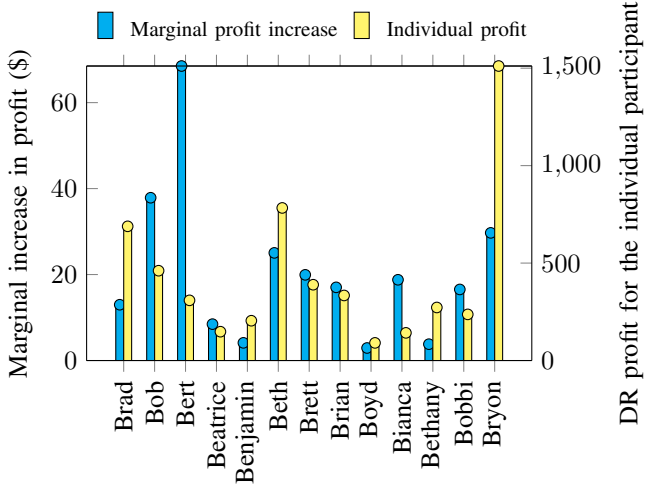
\begin{figure}
\begin{tikzpicture}[
   declare function={
    barW=3pt; 
    barShift=barW/2; 
  }
]
\begin{axis}[
height = 5.5cm,
    ybar,
    bar width=barW, 
    bar shift=-barShift, 
    symbolic x coords={Brad, Bob, Bert, Beatrice, Benjamin,Beth, Brett, Brian, Boyd, Bianca, Bethany, Bobbi, Bryon},
    axis y line*=left,
    ymin=0, ymax=68.52257361,
    ylabel=Marginal increase in profit (\$),
    enlarge x limits=0.1,
    xtick=data, 
    x tick label style={rotate=90,anchor=east},
    legend cell align=left,
        legend columns = 1,
        legend style={
                font=\footnotesize,
                at={(0.3,1.05)},
                anchor=south,
                column sep=1ex,
                draw = none},
        legend image code/.code={
        \draw [#1] (0cm,-0.1cm) rectangle (0.2cm,0.25cm); },
    ]
 
    \addplot[mark=*,
             mark options={xshift=-barShift}, 
             draw=black,
             fill=cyan,
    ] coordinates {
(Brad, 12.98009983)
(Bob, 37.87555584)
(Bert, 68.52257361)
(Beatrice, 8.465609105)
(Benjamin, 4.114863875)
(Beth, 25.0443754)
(Brett, 19.93672324)
(Brian, 17.02437486)
(Boyd, 2.916050149)
(Bianca, 18.78602811)
(Bethany, 3.800928222)
(Bobbi, 16.53643854)
(Bryon, 29.69908783)
    };
    \legend{Marginal profit increase}
\end{axis} 
\begin{axis}[
height = 5.5cm,
    ybar,
    bar width=barW,
    bar shift=barShift,
    symbolic x coords={Brad, Bob, Bert, Beatrice, Benjamin,Beth, Brett, Brian, Boyd, Bianca, Bethany, Bobbi, Bryon},
    axis y line*=right,
    ymin=0, ymax=1511.149291,
    ylabel= DR profit for the individual participant,
    enlarge x limits=0.1,
    xmajorticks = false,
    legend cell align=left,
        legend columns = 1,
        legend style={
                font=\footnotesize,
                at={(0.8,1.05)},
                anchor=south,
                column sep=1ex,
                draw = none},
        legend image code/.code={
        \draw [#1] (0cm,-0.1cm) rectangle (0.2cm,0.25cm); },
]
\addplot[mark=*,
         mark options={xshift=barShift}, 
         fill=yellow!70,
] coordinates {
(Brad, 688.5581454)
(Bob, 460.8411793)
(Bert, 308.7655987)
(Beatrice, 147.9406189)
(Benjamin, 204.2634167)
(Beth, 782.9197488)
(Brett, 388.9390016)
(Brian, 333.8303503)
(Boyd, 90.5373934)
(Bianca, 141.9629198)
(Bethany, 272.1880395)
(Bobbi, 236.733118)
(Bryon, 1511.149291)
    };
    \legend{Individual profit}
\end{axis}
\end{tikzpicture}
\caption{ The marginal increase in profit on aggregating with a building does not depend on its original size} \label{fig:profit-size}
\end{figure}

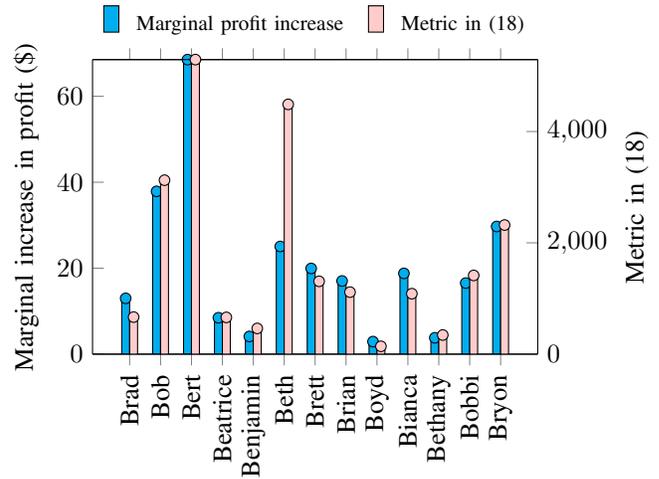
\begin{figure}
\begin{tikzpicture}[
   declare function={
    barW=3pt; 
    barShift=barW/2; 
  }
]
\begin{axis}[
height = 5.5cm, 
    ybar,
    bar width=barW, 
    bar shift=-barShift, 
    symbolic x coords={Brad, Bob, Bert, Beatrice, Benjamin,Beth, Brett, Brian, Boyd, Bianca, Bethany, Bobbi, Bryon},
    axis y line*=left,
    ymin=0, ymax=68.52257361,
    ylabel=Marginal increase in profit (\$),
    enlarge x limits=0.1,
    xtick=data, 
    x tick label style={rotate=90,anchor=east},
    legend cell align=left,
        legend columns = 1,
        legend style={
                font=\footnotesize,
                at={(0.3,1.05)},
                anchor=south,
                column sep=1ex,
                draw = none},
        legend image code/.code={
        \draw [#1] (0cm,-0.1cm) rectangle (0.2cm,0.25cm); },
    ]
 
    \addplot[mark=*,
             mark options={xshift=-barShift}, 
             draw=black,
             fill=cyan,
    ] coordinates {
(Brad, 12.98009983)
(Bob, 37.87555584)
(Bert, 68.52257361)
(Beatrice, 8.465609105)
(Benjamin, 4.114863875)
(Beth, 25.0443754)
(Brett, 19.93672324)
(Brian, 17.02437486)
(Boyd, 2.916050149)
(Bianca, 18.78602811)
(Bethany, 3.800928222)
(Bobbi, 16.53643854)
(Bryon, 29.69908783)
    };
    \legend{Marginal profit increase}
\end{axis} 
\begin{axis}[
height = 5.5cm,
    ybar,
    bar width=barW,
    bar shift=barShift,
    symbolic x coords={Brad, Bob, Bert, Beatrice, Benjamin,Beth, Brett, Brian, Boyd, Bianca, Bethany, Bobbi, Bryon},
    axis y line*=right,
    ymin=0, ymax=5292.1017,
    ylabel=Metric in \eqref{eqn:complementarity},
    enlarge x limits=0.1,
    xmajorticks = false,
    legend cell align=left,
        legend columns = 1,
        legend style={
                font=\footnotesize,
                at={(0.8,1.05)},
                anchor=south,
                column sep=1ex,
                draw = none},
        legend image code/.code={
        \draw [#1] (0cm,-0.1cm) rectangle (0.2cm,0.25cm); },
]
\addplot[mark=*,
         mark options={xshift=barShift}, 
         fill=red!20,
] coordinates {
(Brad, 665.7628513)
(Bob, 3125.496901)
(Bert, 5292.1017)
(Beatrice, 660.4027009)
(Benjamin, 462.0395661)
(Beth, 4487.769054)
(Brett, 1308.311272)
(Brian, 1114.195583)
(Boyd, 140.3659465)
(Bianca, 1085.853092)
(Bethany, 343.6076407)
(Bobbi, 1414.232904)
(Bryon, 2319.880355)
    };
    \legend{Metric in \eqref{eqn:complementarity}}
\end{axis}
\end{tikzpicture}
\caption{The marginal increase in profit when building Benthe forms an aggregation with different buildings is highly correlated with the metric in the complementarity test, with one exception (Beth) \eqref{eqn:complementarity} } \label{fig:sigmas-profit}
\end{figure}

We now validate our test for complementarity, and our results on optimal aggregations. From Fig. \ref{fig:profit-size} we can see that the marginal benefit of aggregation does not depend on the `size' of the individual assets, i.e. we can not assess complementarity by looking at its DR profit as an individual asset. This validates our understanding that estimating asset complementarity is not a straightforward task. 
In Fig. \ref{fig:sigmas-profit} we can see that the complementarity test, i.e. the metric in \eqref{eqn:complementarity} is highly correlated with the marginal benefit of aggregating across different aggregation compositions calculated empirically. This validates our belief that the presented test is an effective measure of complementarity of DR assets.

    
    
    
    
    



\section{Conclusion}

In this paper, we presented a framework for optimizing DR asset participation in incentive based demand response programs. We modeled the DR capability of assets as a random variable, and developed analytic expressions for optimal contract size and profit for a profit maximizing asset. We also explored the effect of variability in capability estimates and risk aversion on both contract size and profit. 
We then explored the marginal benefit of aggregation for such assets, and devised a test for asset complementarity under the assumption that DR curtailment follows a normal distribution. In order to illustrate our results, we proposed a method for estimating the DR capability distributions and applied it to load data.

The usefulness of our work hinges on the ability to estimate probability distributions for DR capability. While we detail a method by which these distributions can be constructed, the accuracy of these models will depend on data availability. Our complementarity test depends on the goodness-of-fit of a normal distribution to the observed data-driven distribution, which may not always be an accurate distribution model for different DR assets.


\bibliographystyle{ieeetr}
\bibliography{main}


\end{document}